\documentclass{PoS}
\usepackage{bbm,amsmath}
\usepackage{cancel,url}
\newcommand{\ben}{\begin{enumerate}}
\newcommand{\een}{\end{enumerate}}
\newcommand{\bit}{\begin{itemize}}
\newcommand{\eit}{\end{itemize}}
\newcommand{\beq}{\begin{equation}}
\newcommand{\eeq}{\end{equation}}
\newcommand{\bea}{\begin{eqnarray}}
\newcommand{\eea}{\end{eqnarray}}
\newcommand{\bean}{\begin{eqnarray*}}
\newcommand{\ean}{\end{eqnarray*}}

\newcommand{\non}{\nonumber}
\newcommand{\tr}{\mbox{Tr}}

\title{Constraint HMC algorithms for gauge-Higgs models}

\ShortTitle{Constraint HMC algorithms for gauge-Higgs models}

\author{\speaker{Roman H\"ollwieser}\thanks{Supported by DFG under SFB/TRR55.}\\
Department of Physics, Fakult\"at für Mathematik und Naturwissenschaften,\\ 
Bergische Universit\"at Wuppertal, Gau{\ss}stra{\ss}e 20, 42119 Wuppertal, Germany\\
        E-mail: \email{hoellwieser@uni-wuppertal.de}}

\author{Francesco Knechtli\\
Department of Physics, Fakult\"at für Mathematik und Naturwissenschaften,\\ 
Bergische Universit\"at Wuppertal, Gau{\ss}stra{\ss}e 20, 42119 Wuppertal, Germany\\
        E-mail: \email{knechtli@physik.uni-wuppertal.de}}

\abstract{We present the construction of constraint HMC algorithms for
gauge-Higgs models in order to measure the effective Higgs potential.
In particular we focus on SU(2) Gauge-Higgs Unification models in five dimensions. 
Previous simulations have identified regions in the
Higgs phase of these models which have properties of 4D adjoint or Abelian Higgs
models. We want to test this relationship by comparing the effective
potentials in five and four dimensions.}

\FullConference{The 36th Annual International Symposium on Lattice Field Theory - LATTICE2018\\
		22-28 July, 2018\\
		Michigan State University, East Lansing, Michigan, USA.}

\begin{document}

\section{INTRODUCTION}

The discovery in 2012 of a scalar particle around $125$ GeV has all but confirmed the existence of the Higgs mechanism, which renders the standard model of particle physics complete. However, the origin of the potential responsible for Spontaneous Symmetry Breaking (SSB), which leads to the Brout-Englert-Higgs (BEH) mechanism \cite{Englert:1964et,Higgs:1964ia} %,Higgs:1964pj} 
is, as of yet, unknown, and the arbitrary nature of the required fine tuning of parameters, the so-called hierarchy problem, suggests that a more fundamental process is at work. %Moreover, the recent observation of a diphoton excess in the scalar channel at about $750$ GeV at the ATLAS and CMS experiments \cite{CMS:2015dxe,ATLAS:2015hp}, find no explanation within the standard model, bringing forth the need for a more general theory.

A class of extensions to the Standard Model aimed at addressing these puzzles by the use of extra dimensions come under the heading of Gauge-Higgs Unification (GHU) \cite{Manton:1979kb,Fairlie:1979at,Hosotani:1983vn}. 
In these models, the Higgs field originates from the extra-dimensional components of the gauge field and gives rise to massive gauge bosons in the regular four dimensions. In this study, we consider the simplest case of one extra dimension where, due to the higher dimensional gauge invariance, the Higgs potential remains zero at tree level and is generated only through quantum effects \cite{Coleman:1973jx}. The question is where and how dimensional reduction from five to four dimensions occurs. Possible mechanisms, which have been proposed are compactification or localization. 

In this work we are aiming at the study of the dimensional reduction of the 5D GHU models with torus and orbifold boundary conditions and their connection to the 4D adjoint resp. Abelian-Higgs model by analyzing the effective potentials in both models. In particular, we want to see whether this effective potential reproduces the form of the SM potential, {\it i.e.}:
\beq
V_{eff}(H)=-\mu^2HH^\dagger+\lambda(HH^\dagger)^2,
\eeq
where $H=(H_0,H^+)$ is the SM Higgs doublet and $\mu^2$ and $\lambda$ are the well-known Higgs mass and Higgs self-coupling parameters. Notice that, in order to have spontaneous symmetry breaking of the electroweak symmetry, $\mu^2$ must be negative and $\lambda$ must be positive to have a well-defined energy minimum, giving the well-known Mexican hat potential. In addition these parameters should reproduce the SM relation $m_H^2=2\lambda\nu^2=2\mu^2$, where $m_H$ is the Higgs mass and $\nu$ is the vacuum expectation value (vev). The goal is to measure the so-called constraint effective potential in lattice simulations, which in the infinite volume limit corresponds to the conventional effective potential~\cite{Fukuda:1974ey,ORaifeartaigh:1986axd}. 
 Therefore we need to formulate constraint Hamiltonians for the various models, {\it i.e.} the energy functional including constraint conditions on the corresponding Higgs fields, and derive constraint equations of motion. In these proceedings we discuss the implementation of the latter in constraint Hybrid Monte Carlo algorithms for the 4D Abelian-Higgs and the 5D models with torus and orbifold boundary conditions.~%, which turns out to be a non-trivial task, since the standard leap-frog algorithm fails in this approach. 
 We start with a short introduction of the 5D GHU model before discussing the constraint HMC algorithms in section~\ref{sec:chmc}. We conclude with final remarks and an outlook to other interesting applications of these constraint algorithms to measure effective potentials, {\it e.g.}, in finite temperature QCD.

\section{THE 5D ORBIFOLD MODEL}\label{sec:5dmodel}

\iffalse
\begin{figure}[h]
\vspace{-5mm}
\centering
a)
\includegraphics[width=0.41\textwidth]{latorb.pdf}
b)\includegraphics[width=0.52\textwidth]{phasediagram.pdf}
\caption{a) A sketch of the orbifold lattice: the $SU(2)$ links are depicted in blue and the $U(1)$ links on the boundaries in red, the magenta links sticking out of the boundary are the so-called hybrid links, which gauge-transform as $SU(2)$ on one end and as $U(1)$ on the other. b) The phase diagram for $N_5 = 4$ in the region of the Higgs-hybrid phase transition. The points show the location of a first-order phase transition with the width of the corresponding hystereses in red and blue, while the dashed orange line represents $\gamma = 1$.} 
\label{f_orbisketch}
\end{figure}
\fi

%The theory represented in Fig.~\ref{f_orbisketch}a is 
Our prototype to study the Brout-Englert-Higgs (BEH) mechanism, 
i.e. the spontaneous symmetry breaking (SSB) of the, in this case, $U(1)$ 
gauge symmetry on the boundary to nothing is given by the anisotropic
lattice action~\cite{Irges:2004gy}
\begin{equation}\label{eq:orbiaction}
S_W^{orb} = \frac{\beta_4}{2}\sum_{\mu,\nu<4} w\,\tr\{1-U_{\mu\nu}\} +
\frac{\beta_5}{2}\sum_{\mu<4}\tr\{1-U_{\mu5}\} \,,
\end{equation}
where $\beta_4$ and $\beta_5$ are the gauge couplings associated with plaquettes
spanning the standard four dimensions ($U_{\mu\nu}$) and the fifth dimension ($U_{\mu5}$)
respectively. In the sums of Eq.~\ref{eq:orbiaction} plaquettes are counted with
one orientation only. The weight $w$ associated with plaquettes $P_4$ on the
boundaries takes a value $w=1/2$ and it is $w=1$ otherwise. The boundary links are in the gauge group $U(1)$ and all other links are in $SU(2)$.
The anisotropy is $\gamma=\sqrt{\beta_5/\beta_4}$ and in the classical limit
$\gamma=a_4/a_5$, where $a_4$ denotes the lattice spacing in the usual
four dimensions and $a_5$ denotes the lattice spacing in the extra dimension.
The theory is defined on the interval $I=\{n_\mu,0\le n_5\le N_5\}$, where
$(n_\mu,n_5)$, $\mu=0,1,2,3$ are the integer coordinates of the points. 

The Higgs field is constructed from the Polyakov lines $P$ in the extra dimension, {\it i.e.}, the product of link variables along the $5^{th}$ direction times the orbifold boundary element $g=-i\sigma_3$ and the complex conjugate of the whole thing in order to get a closed loop: 
\beq
P(x)=\prod_{n_5=0}^{N_5-1}[U_5(x,n_5a_5)]\,g\prod_{n_5=N_5-1}^0[U^\dagger_5(x,n_5a_5)]\,g^\dagger.\label{eq:pol}
\eeq

We found that it exhibits a phase with Spontaneous Symmetry Breaking (SSB) with a massive gauge boson. 
Moreover on the orbifold boundaries we observed dimensional reduction from five to four dimensions, which suggests that there is a localization mechanism for the gauge field. These results, which are favorably pointing towards the suitability of this theory for describing the electro-weak sector of the standard model, are reported in \cite{Alberti:2015pha}.

\section{THE CONSTRAINT HMC ALGORITHM}\label{sec:chmc}

One can calculate the exact effective potential non-perturbatively, using lattice simulations. This was first shown in the pure Higgs theory by Kuti and Shen \cite{Kuti:1987bs}, via simulating the constraint path integral,
\beq
e^{-\Omega U_\Omega(\Phi)}=\int\mathcal{D}\phi\delta(\frac{1}{\Omega}\sum_x\phi(x)-\Phi) e^{-S[\phi]}\label{eq:dUint}
\eeq
where $\Omega$ is the total volume and the average of the Higgs field $\phi(x)$ is forced to fluctuate around a fixed value $\Phi$. During the constraint simulations we measure the derivative of the constraint effective potential $U_\Omega$ with respect to the constraint field $\Phi$ and a separate lattice simulation has to be run for every value of $\Phi$. This method is computationally expensive, but determines the effective potential with greater accuracy than fitting a distribution $P(\Phi)$. In the infinite volume limit the constraint potential gives the effective potential $U_\Omega(\Phi)\overset{\Omega\rightarrow\infty}{\rightarrow}U_{eff}(\Phi)$~\cite{Fukuda:1974ey,ORaifeartaigh:1986axd}. 
 The simulations are performed using Hybrid Monte Carlo methods implementing constraint Hamiltonian equations of motion. The latter can be derived by rewriting Eq.~\ref{eq:dUint} in terms of the constraint Hamiltonian
\bea
H[\phi,\pi]&=&S[\phi]+\frac{1}{2}\sum_x\pi^2(x)+\xi\biggl(\frac{1}{\Omega}\sum_x\phi(x)-\Phi\biggr)+\ldots\label{eq:cham}\\
e^{-\Omega U_\Omega(\Phi)}&=&\int\mathcal{D}\phi\delta(\frac{1}{\Omega}\sum_x\phi(x)-\Phi) e^{-S[\phi]}=\int\mathcal{D}\phi\mathcal{D}\pi e^{-H[\phi,\pi]}
\eea
where the fictitious momentum variables $\pi(x)$ are introduced and the Lagrange multiplier $\xi$ ensures that the Higgs field $\phi(x)$ fluctuates around a fixed average value $\Phi$. The dots indicate that there might be further Lagrange multipliers in the Hamiltonian, incorporating the so-called hidden constraints, {\it i.e.} time derivatives of the constraint condition. In the case of a constraint that is linear in the underlying fields, {\it i.e.}, if the field Higgs field $\phi(x)$ is a real scalar field, the hidden constraint only depends on the momenta $\pi(x)$ and a standard leap-frog algorithm can be applied. If the constraint is applied to non-composite fields however, we get additional conditions of the form $\sum_x\dot\phi(x)$, depending on $\pi(x)$ and $\phi(x)$, which during a standard leap-frog trajectory are never defined at the same integration time and therefore the hidden constrain cannot be evaluated. This is also the case for $SU(N)$ fields, where the change of the fields in HMC algorithms is not given by an additive but an exponential term proportional to the momenta $\pi(x)$, as in sections~\ref{sec:chmctor} and~\ref{sec:chmcorb}. 

In order to implement the constraint equations of motion for these special cases we use an extension of the so-called Rattle algorithm to general Hamiltonians for constraint systems, the Newton-St\"ormer-Verlet-leapfrog method~\cite{hairer:2002gni} %,hairer:2003gni}. 
These generalized leap-frog algorithms have an additional half integration step to get $\pi_{n+1/2}$ to $\pi_{n+1}$ in order to have the momentum $\pi$ at the same integration time as the field variable. We use the index $n$ for the molecular dynamics time steps $nh$, where $h$ is the integration step size. This allows us to apply the so-called hidden constraint, which is the first derivative with respect to (integration) time of the constraint condition and involves both fields. We successfully implemented the algorithms and numerically tested their time-reversibility and symplecticity. In the following we summarize the new algorithms for the various models. 

\subsection{4D Abelian-Higgs model}\label{sec:chmcabel}
In order to respect gauge invariance the 4D Abelian-Higgs model the constraint condition reads $\dfrac{1}{\Omega}\sum_x\phi_{n}^\dagger(x)\phi_{n}(x)=\Phi$, the constraint HMC algorithm for the scalar field is given by
\bean
\pi_{n+1/2}&=&\pi_n-\dfrac{h}{2}\left(\dfrac{\partial S}{\partial\phi_n}+\dfrac{2\phi_n\lambda_n}{\Omega}\right)\;,
\quad \phi_{n+1}=\phi_n+h\pi_{n+1/2}\\
\lambda_n&=&\dfrac{\Omega}{h^2}-\sum_x\dfrac{\phi_n}{2\Phi}\dfrac{\partial S}{\partial\phi_n}\pm\sqrt{\dfrac{\Omega^2}{h^4}+\bigg(\sum_x\dfrac{\phi_n}{2\Phi}\dfrac{\partial S}{\partial\phi_n}\bigg)^2-\dfrac{\Omega}{\Phi}\sum_x\left(\dfrac{\pi_n^2}{h^2}-\dfrac{\pi_n}{h}\dfrac{\partial S}{\partial\phi_n}+\dfrac{1}{4}\bigg(\dfrac{\partial S}{\partial\phi_n}\bigg)^2\right)}\\
\pi_{n+1}&=&\pi_{n+1/2}-\dfrac{h}{2}\left(\dfrac{\partial S}{\partial\phi_{n+1}}+\dfrac{2\phi_{n+1}\mu_n}{\Omega}\right),\qquad
\mu_n=\sum_x\bigg(\dfrac{\phi_{n+1}\pi_{n+1/2}}{h\Phi}-\dfrac{\phi_{n+1}}{2\Phi}\dfrac{\partial S}{\partial\phi_{n+1}}\bigg),
\ean
where $\phi_n\equiv\phi_n(x)$ at MD time $nh$, $\Omega$ the total volume and $S$ the Abelian-Higgs action
\bea
S[U,\phi]&=&S_g[U]+S_\phi[U,\phi], \quad S_g[U]=\beta\sum_x\sum_{\mu<\nu}\left\{1-\mbox{Re}U_{\mu\nu}(x)\right\}\\
S_\phi[U,\phi] &=& \sum_x
|\phi(x)|^2-2\kappa\sum_\mu\mbox{Re}\left\{\phi^\dagger(x)[U_\mu(x)]^q\phi(x+a\hat\mu)\right\}+\lambda(|\phi(x)|^2-1)^2\label{eq:Sabel}
\eea
with the $U(1)$ gauge links $U_\mu(x)$ and $U_{\mu\nu}(x)=U_\mu(x)U_\nu(x+a\hat\mu)U_\mu^\dagger(x+a\hat\nu)U_\nu^\dagger(x)$ the standard plaquettes. $a$ is the lattice spacing and we use $q=1$. During numerical simulations it turns out that only the $-$ sign in front of the square root in $\lambda_n$ fulfills the constraint condition. 

Note, we can also write the action~\ref{eq:Sabel} in unitary gauge using the variable transformation proposed in \cite{Montvay:1994cy} p.322ff, $\phi(x)=\rho(x)\exp{i\varphi(x)}\;\Rightarrow\;\phi_1=\rho\cos\varphi,\;\phi_2=\rho\sin\varphi$:
\bean
S_\phi[U,\rho,\varphi]&=&\sum_x\rho_x^2+\lambda(\rho_x^2-1)^2-2\kappa\rho_x\sum_\mu\rho_{x+\hat\mu}\underbrace{e^{-\varphi_{x+\hat\mu}}U_{x,\mu}e^{i\varphi_x}}_{=V_{x,\mu}}=S_\rho[V,\rho],
\ean
which allows us to rewrite the constraint Hamiltonian and equations of motion as
\bean
H[V,\rho]&=&S_\rho[V,\rho]+\dfrac{1}{2}\sum_x\pi(x)^2+\mu\left(\dfrac{1}{\Omega}\sum_x\rho(x)-\Phi\right)+\sigma\left(\dfrac{1}{\Omega}\sum_x\pi(x)\right)\\
\dot\rho(x,t)&=&\dfrac{\partial H}{\partial\pi(x,t)}=\pi(x,t)+\dfrac{\sigma}{\Omega};\quad\dot\pi(x,t)=-\dfrac{\partial H}{\partial\rho(x,t)}=-\dfrac{\partial S_\rho}{\partial\rho(x,t)}-\dfrac{\mu}{\Omega}\\
\sum_x\dot\rho(x)&=&0\Rightarrow\sigma=-\sum_x\pi(x)=0;\quad\sum_x\dot\pi(x)=0\Rightarrow\mu=-\sum_x\dfrac{\partial S_\rho}{\partial\rho(x)}
\ean
and use the standard leap-frog algorithm as shown in~\cite{Fodor:2007fn} for a Higgs-Yukawa theory with $N_f$ fermions. In order to guarantee that the hidden constraint is fulfilled by the algorithm, one has to initialize the (random) fictitious momenta $\pi(x)$ in each trajectory accordingly, {\it i.e.}, with respect to $\sum_x\pi(x)=0$. During the constraint simulations we measure the derivative of the effective potential
\beq
\dfrac{dU_\Omega}{d\Phi}=2\Phi+4\lambda\bigg\langle\dfrac{1}{\Omega}\sum_x(\rho(x)^2-1)\rho(x)\bigg\rangle_\Phi+2\kappa\bigg\langle\dfrac{1}{\Omega}\sum_{x,\mu}(\rho(x-\hat\mu)V_\mu(x-\hat\mu)+\rho(x+\hat\mu))V_\mu(x)\bigg\rangle_\Phi\label{eq:ceffP}
\eeq
where $\langle\ldots\rangle_\Phi$ means the expectation value at fixed $\Phi=\Omega^{-1}\sum_x\rho(x)$. Results are presented in Fig.1 for a simulation in the Higgs phase. The potential has indeed the Mexican hat form.

\begin{figure}
\centering
\includegraphics[width=0.75\linewidth]{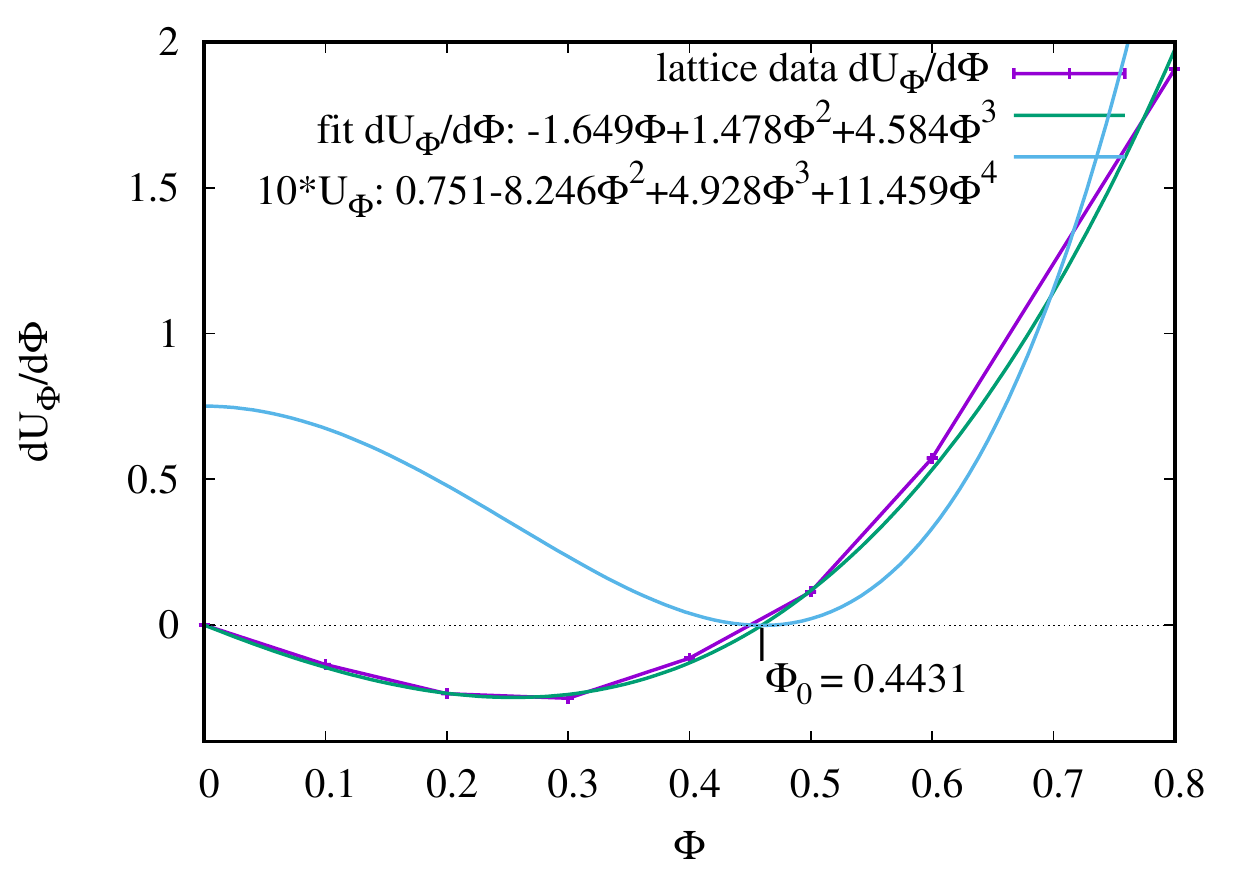}
\vspace{-3mm}
\caption{The constraint effective potential (\ref{eq:ceffP}) of the Abelian-Higgs model in unitary gauge for $\lambda=1, \kappa=0.3$ and $\beta=1$. We fit the data with $2c_1\Phi+3c_2\Phi^2+4c_3\Phi^3$~\cite{Irges:2017ztc} and plot its integral using an integration constant to shift the potential minimum to zero and seizing the potential by a factor 10 for visibility. The unconstrained vacuum expectation value (vev) of the field $\langle\rho\rangle$ agrees with the potential minimum at $\Phi_0$.}
\end{figure}

\subsection{5D Torus GHU model}\label{sec:chmctor}

In the 5D $SU(2)$ gauge theory we have to fix the average of the extra dimensional Polyakov lines, which represent the Higgs field. This is done via gauge transforming the $U_5$ links to either boundary, leaving a single "boundary" link $V_5(x)=\prod_{n_5=0}^{N_5-1}[U_5(x,n_5a_5)]$, representing the Polyakov loop, and applying the constraint $\dfrac{1}{2\Omega}\sum_x\tr V_5(x)=\Phi$ during the constraint HMC algorithm
\bean
\pi_{n+1/2}&=&\pi_n-\dfrac{h}{2}\left(\dfrac{\partial S}{\partial V_n}-\dfrac{\lambda_n}{8\Omega}\tr[\sigma_i V_n]\sigma^i\right),\quad V_{n+1}=e^{h\pi_{n+1/2}}V_n\\
\dfrac{\lambda_n}{8\Omega}&=&\left(\sum_x\tr\dfrac{\partial S}{\partial V_n}V_n-\sum_x\tr\pi_n^2V_n\right)/\sum_x\tr[\sigma_i V_n]\tr[\sigma_i V_n]\\
\pi_{n+1}&=&\pi_{n+1/2}-\dfrac{h}{2}\left(\dfrac{\partial S}{\partial V_{n+1}}-\dfrac{\mu_n}{8\Omega}\tr[\sigma_i V_{n+1}]\sigma^i\right)\\
\dfrac{\mu_n}{8\Omega}&=&\left(\sum_x\tr\dfrac{\partial S}{\partial V_{n+1}}V_{n+1}-2\sum_x\tr\pi_{n+1/2}V_{n+1}/h\right)/\sum_x\tr[\sigma_i V_{n+1}]\tr[\sigma_i V_{n+1}]\\
\ean
where $V_n\equiv V_5(x)$ at MD time $nh$, $\Omega$ the 4D volume and the action $S$ given by Eq. (\ref{eq:orbiaction}) with $w=1$ everywhere and links resp. plaquettes periodically connecting around all five boundaries.

\subsection{5D Orbifold gauge-Higgs model}\label{sec:chmcorb}

In the case of orbifold boundary conditions, the Polyakov line (\ref{eq:pol}) in axial gauge reads $P(x)=V_5(x)\sigma_3V^\dagger_5(x)\sigma_3$ and the constraint condition is $\dfrac{1}{2\Omega}\sum_x\tr (V_5(x)\sigma_3V_5(x)^\dagger\sigma_3)=\Phi$. The constraint HMC algorithm is given by
\bean
\pi_{n+1/2}&=&\pi_n-\dfrac{h}{2}\left(\dfrac{\partial S}{\partial V_n}-\dfrac{\lambda_n}{8\Omega}\tr([\sigma_3,\sigma_i] V_n\sigma_3V_n^\dagger)\sigma^i\right),\quad V_{n+1}=e^{h\pi_{n+1/2}}V_n\\
\dfrac{\lambda_n}{8\Omega}&=&\dfrac{\sum_x\left(2\tr(\pi_nV_n\sigma_3V_n^\dagger\pi_n\sigma_3)+\tr([\sigma_3,\partial S/\partial V_n]V_n\sigma_3V_n^\dagger)-2\tr(\pi_n^2V_n\sigma_3V_n^\dagger\sigma_3)\right)}{\sum_x\tr([\sigma_3,\sigma_i]V_n\sigma_3V_n^\dagger)\tr([\sigma_3,\sigma_i]V_n\sigma_3V_n^\dagger)}\\
\pi_{n+1}&=&\pi_{n+1/2}-\dfrac{h}{2}\left(\dfrac{\partial S}{\partial V_{n+1}}-\dfrac{\mu_n}{8\Omega}\tr([\sigma_3,\sigma_i] V_{n+1}\sigma_3V_{n+1}^\dagger)\sigma^i\right)\\
\dfrac{\mu_n}{8\Omega}&=&\sum_x\bigg(\,\tr([\sigma_3,\sigma_i] V_{n+1}\sigma_3V_{n+1}^\dagger)\tr(\sigma^i\partial S/\partial V_{n+1})-2\tr([\sigma_3,\sigma_i] V_{n+1}\sigma_3V_{n+1}^\dagger)\cdot\\&&\qquad\tr(\sigma^i\pi_{n+1/2})/h\bigg)/\sum_x\tr([\sigma_3,\sigma_i] V_{n+1}\sigma_3V_{n+1}^\dagger)\tr([\sigma_3,\sigma_i] V_{n+1}\sigma_3V_{n+1}^\dagger)
\ean

\section{CONCLUSIONS AND OUTLOOK}

We successfully implemented the constraint HMC algorithms for gauge-Higgs models outlined in section~\ref{sec:chmc} in three particular models, namely the 4D Abelian-Higgs and a 5D $SU(2)$ GHU model with torus and orbifold boundary conditions. In the Higgs phase of the 4D Abelian-Higgs model we observe a Mexican hat form for the constraint effective potential. The algorithm in section~\ref{sec:chmctor} can be adopted wholesale to $SU(2)$ gauge theory in four dimensions, where it may turn out very useful for finite temperature studies of (constraint) effective Polyakov line actions. For more details and an extensive analysis of results on the constraint effective potential we refer to an upcoming publication~\cite{Gunther:2018}.

% Acknowledgement
\section{ACKNOWLEDGMENTS}
We thank Michael G\"unther and Julius Kuti for helpful discussions. We gratefully acknowledge the Gauss Center for Supercomputing (GCS) for providing computer time at the supercomputers JURECA/JUWELS at the Juelich Supercomputing Centre (JSC) under GCS/NIC project ID HWU24. This work is supported by the Deutsche Forschungsgemeinschaft in the SFB/TRR55.

\bibliographystyle{utphys}
\bibliography{PoSlatt18}

\end{document}